\documentclass[aps,prb,twocolumn,superscriptaddress,a4paper]{revtex4-1}

\usepackage{amsmath}
\usepackage[dvipdfmx]{graphicx}
\usepackage{adjustbox}
\usepackage{bm}
\usepackage{color}
\usepackage{braket}
\usepackage{standalone}
\usepackage{multirow}
\usepackage{tikz}
\usepackage{mathrsfs}
\usepackage{amssymb}
\usepackage{bbold}
\usepackage[colorlinks,bookmarks=true,citecolor=blue,linkcolor=red,urlcolor=blue]{hyperref}

\newcommand{\1}{\mbox{1}\hspace{-0.25em}\mbox{l}}

\begin{document}
\title{Phase Transitions and Generalized Biorthogonal Polarization\\ in Non-Hermitian Systems }

\author{Elisabet Edvardsson}
\affiliation{Department of Physics, Stockholm University, AlbaNova University Center, 106 91 Stockholm, Sweden}
\author{Flore K. Kunst}
\affiliation{Max-Planck-Institut f\"{u}r Quantenoptik, Hans-Kopfermann-Stra{\ss}e 1${\rm ,}$ 85748 Garching${\rm ,}$ Germany}
\author{Tsuneya Yoshida}
 \affiliation{Department of Physics, University of Tsukuba, Ibaraki 305-8571, Japan}
\author{Emil J. Bergholtz}
\affiliation{Department of Physics, Stockholm University, AlbaNova University Center, 106 91 Stockholm, Sweden}
\date{\today}

\begin{abstract}
Non-Hermitian (NH) Hamiltonians can be used to describe dissipative systems, and are currently intensively studied in the context of topology. A salient difference between Hermitian and NH models is the breakdown of the conventional bulk-boundary correspondence invalidating the use of topological invariants computed from the Bloch bands to characterize boundary modes in generic NH systems. One way to overcome this difficulty is to use the framework of biorthogonal quantum mechanics to define a biorthogonal polarization, which functions as a real-space invariant signaling the presence of boundary states. Here, we generalize the concept of the biorthogonal polarization beyond the previous results to systems with any number of boundary modes, and show that it is invariant under basis transformations as well as local unitary transformations. Additionally, we propose a generalization of a perviously-developed method with which to find all the bulk states of system with open boundaries to NH models. Using the exact solutions in combination with variational states, we elucidate genuinely NH aspects of the interplay between bulk and boundary at the phase transitions.
\end{abstract}

\maketitle

\section{Introduction}
One of the fundamental postulates of quantum mechanics is the assumption that observables are described by Hermitian operators, which ensures realness of the measured eigenvalues. This, however, fails to take into account that in reality systems typically interact with the environment giving rise to dissipation and other non-equilibrium phenomena. An effective approach to describe such open systems is by making use of \emph{non-Hermitian} (NH) operators. The study of NH Hamiltonians has in the past years become increasingly popular and finds applications in classical systems, e.g., in optics \cite{SzReBaSe2011,ReBeMiOnChPe2012,Wi2014,HoHaWiGaElChKh2017,FeElGe2017,SoAl2017,HaBaLuReChKhChSe2018,BaWiHaPaReSeChKh2018,ZhMiTeMaElScFe2018,KrBiMaHeThSz2019,OzRoNoYa2019}, electric circuits \cite{NiOwSoScSi2015,AlGlJi2015,LeImBeBaBrMoKiTh2018,Ez2019,HeHoImAbKiMoLeSzGrTh2019,HoHeScSaBrGrKiWoVoKaLeBiThNe2019,Yoshida_MSkin2020}, 
and topological mechanical metamaterials \cite{NaKlReViTuIr2015,BrLoLeCo2019,GhBrWeCo2019,Yoshida_nHMech2019}, but also in quantum systems, such as quasi-particles with finite lifetimes in heavy-fermion systems \cite{Kozii2017, Yoshida2018,Yoshida_nHQPRev2020}, and material junctions \cite{Bergholtz2019nonHermitian}. 
Recently, there has been an increasing focus on studying the topological properties of such NH systems \cite{BeBuKu2019}, which have been studied both theoretically \cite{KaBeSa2019,LeLiGo2019,EdKuBe2019,CaStBuBe2019,BuCaKuBe2019,Yoshida2019,CaBe2018,KochBBC,GoAsKaTaHiUe2018,ZhLe2019,StRoArBuBe2019,LeBlHuChNo2017,ShZhFu2018,LuPi2019,KaShUeSa2019,LuZh2019,EsSaHaKo2011,YaWa2018,KuEdBuBe2018,AlVaTo2018,Xi2018,Le2016,XiWaWaTo2016,KuDw2019,Fanny,Yoshida_nHFQH19,Yoshida_nHFQH20} and experimentally \cite{WaChJoSo2009,ZeRePlLuNoRuSeSz2015,PeOzLiChKrYiWiRoYa2016,WeKrPlLuNoMaSeReSz2017,ChOzZhWiYa2017,CeHuWaChChRe2019,XiDeWaZhWaYiXu2019,HeQiYeCaFaKeZhLi2018,ZhPeYoHsNeFuJoSoZh2018,ZhLeLiZh2019,PoBeKuMoSc2015}. 

Alleviating the Hermiticity condition may introduce effects that at first glance seem surprising or unintuitive, such as the possible breakdown of the conventional bulk-boundary correspondence (BBC) \cite{YaWa2018,EdKuBe2019,KuEdBuBe2018,AlVaTo2018,Xi2018,Le2016,XiWaWaTo2016, KochBBC,KuDw2019}. This phenomenon is accompanied by the so-called NH skin effect, which refers to the piling up of bulk states at the boundaries \cite{YaWa2018}, as well as the appearance of exceptional points (EPs), which are degeneracies at which the geometric multiplicity is smaller than the algebraic multiplicity, whose order scales with system size \cite{KuDw2019}. The breakdown of the conventional BBC as well as the emergence of the NH skin effect has been experimentally verified in mechanical systems \cite{GhBrWeCo2019,BrLoLeCo2019}, topoelectrical circuits \cite{HeHoImAbKiMoLeSzGrTh2019} and optical \cite{XiDeWaZhWaYiXu2019} systems. This phenomenology has also been suggested to be of practical use in sensors whose sensitivity increases exponentially with the size of the system \cite{NHsensor}.

Crucially, when the conventional BBC is broken, it is no longer possible to directly use topological invariants derived from the Bloch Hamiltonian to characterize the topological phase of the system, and to predict the presence of boundary states.
In Refs.~\onlinecite{KuEdBuBe2018, EdKuBe2019} several of the authors of this work proposed an alternative BBC to remedy this breakdown called the \emph{biorthogonal bulk-boundary correspondence}, which finds its basis in biorthogonal quantum mechanics \cite{Br2013}. Explicitly using that the left and right eigenstates of an NH Hamiltonian are generally different and non-orthogonal, the biorthogonal BBC considers the combination of these eigenstates to accurately predict the localization of boundary modes as well as gap closings in the open-boundary-condition (OBC) spectrum \cite{KuEdBuBe2018}.

Indeed, one of the central results of Ref.~\onlinecite{KuEdBuBe2018} is the introduction of the \emph{biorthogonal polarization}
\begin{equation}
\mathcal{P} = 1-\lim_{N\rightarrow\infty}\frac{1}{N}\braket{\psi_L|\sum_n n\hat{\Pi}_n|\psi_R}, \label{eq:original_biorth_pol}
\end{equation}
where $\hat{\Pi}_n$ is the projection operator onto the $n$th unit cell of the lattice with OBC, $N$ is the total number of unit cells, and $\psi_L$ and $\psi_R$ are the left and right eigenstates, respectively, of the boundary mode.
This predicts the presence ($\mathcal{P}=1$) or absence ($\mathcal{P}=0$) of a boundary mode on each boundary in quasi one-dimensional systems, i.e., systems with OBC in one direction, and can thus be interpreted as a real-space invariant. In this paper, we generalize this quantity to quasi one-dimensional systems with \emph{any} number of boundary modes, and show with an example that its value corresponds exactly to the number of boundary modes on the boundaries. Additionally, we show that the polarization is invariant under gauge transformations as well as unitary transformations that are local thus corroborating the invariance of the polarization.

We also present a study of the OBC properties of the anisotropic Su-Schrieffer-Heeger (SSH) chain [cf. Fig.~\ref{fig:Hamiltonian}(b)] by studying the gap closings as well as analytical solutions for all the bulk states. By making use of analytical results from Refs.~\onlinecite{KuEdBuBe2018, KuDw2019} for the periodic-boundary-condition (PBC) and OBC cases, we study the behavior of the band-gap closing in the PBC and OBC spectrum and find that they scale differently with system size. Additionally, we show that the method in Ref.~\onlinecite{KuMiBe2019} for finding all bulk-state solutions analytically can be extended to the NH realm. Whereas in Hermitian systems this method relies on a \emph{spectral mirror symmetry} in the Bloch spectrum relating the eigenvalues at $k$ to the eigenvalues at $-k$, here we find that this symmetry only needs to be present in the OBC spectrum, i.e., $E_\textrm{OBC}(k) = E_\textrm{OBC}(-k)$ whereas it may be absent in the Bloch spectrum. By taking a closer look at these solutions at the gap closings, we can show that they are equivalent to the boundary states up to a twist thus proving that the gap indeed disappears.

This paper is organized as follows: In Sect.~\ref{sect:biorthogonal_polarization}, we introduce the generalized biorthogonal polarization and discuss its properties. This is followed by a thorough study of the anistropic SSH chain in Sect.~\ref{sect:bulk_states_gap_closings}. Lastly, we conclude with a discussion in Sect.~\ref{sect:discussion}.

\section{The biorthogonal polarization} \label{sect:biorthogonal_polarization}

In this section, we define, generalize and discuss the properties of the biorthogonal polarization, which was originally introduced in Ref.~\onlinecite{KuEdBuBe2018} for quasi-one-dimensional models with a maximum of one boundary state on each boundary [cf. Eq.~(\ref{eq:original_biorth_pol})]. We note that throughout this section we assume models with OBCs, where the boundaries have codimension one.

\subsection{Basic properties of the biorthogonal polarization}
We define the following generalized biorthogonal polarization operator
\begin{equation}
	\hat{P} = {\1} -\lim_{N\rightarrow\infty}\frac{1}{N}\sum_{n=1}^{N}n\hat{\Pi}_n,
\end{equation}
where $N$ is the total number of unit cells in the system, and $\hat{\Pi}_n = \sum_{m}\ket{e_{nm}}\bra{e_{nm}}$ with $\ket{e_{nm}} \equiv c^\dagger_{nm} \ket{0}$ is a projection operator that projects onto the $n$th unit cell with $m$ labelling the internal degrees of freedom inside the unit cell $n$.
From this we define the biorthogonal polarization $\mathcal{P}$ as 
\begin{equation}
	\mathcal{P} = \mathrm{tr}\left[P_{\alpha\beta}\right],
\end{equation}
where the trace is over the matrix $P_{\alpha\beta}$ with matrix elements
\begin{equation}
	P_{\alpha\beta} = \braket{\psi_{\alpha L}|\hat{P}|\psi_{\beta R}},
\end{equation}
where $\ket{\psi_{\alpha R/L}}$ labels the right/left boundary modes. For a system with $M$ gapless edge modes, we thus find
\begin{equation}
	\mathcal{P} = M-\lim_{N\rightarrow\infty}\frac{1}{N}\sum_{\alpha= 1}^{M}\braket{\psi_{\alpha L}|\left(\sum_{n=1}^{N}n\hat{\Pi}_n\right)|\psi_{\alpha R}}, \label{eq:general_biorth_pol}
\end{equation}
where we see immediately that we retrieve Eq.~(\ref{eq:original_biorth_pol}) for $M=1$.

The biorthogonal polarization $\mathcal{P}$ takes integer values $\mathcal{P} \in \mathbb{Z}$. 
This was explained in Ref.~\onlinecite{KuEdBuBe2018}, and we summarize the argument here for the sake of completion. 
Assume a lattice model with a broken unit cell at one of the boundaries, such that each boundary state contained in $M$ always exists regardless of the choice of parameters, 
where the parameter choices determine the boundary on which the state is localized. 
Additionally, assume that the boundary states are chosen in such a way that they are biorthogonal to each other, i.e.,
\begin{equation}
\braket{\psi_{\alpha L}|\psi_{\beta R}} = \delta_{\alpha\beta} \braket{\psi_{\alpha L}|\psi_{\alpha R}} = \delta_{\alpha\beta}.
\end{equation}
Let us focus on what happens when a boundary state is localized to unit cell $n=1$. In this case, the limit in Eq.~(\ref{eq:general_biorth_pol}) goes to zero. Similarly, if the state is localized to unit cell $n=N$, the limit goes to one. Therefore, each localized state contributes either a zero or a one to $\mathcal{P}$, and $\mathcal{P}$ must thus be quantized.

Another consequence of the above explanation is that the biorthogonal polarization $\mathcal{P}$ of a system with a broken unit cell tells us how many of the boundary states in the system are localized to the boundary at $n=1$ (while $M - \mathcal{P}$ tells us how many boundary states are localized at $n=N$).
Crucially, $\mathcal{P}$ is also a relevant quantity for systems without a broken unit cell. In this case, one can think of the lattice as having a mirror symmetry up to local permutations of the internal degrees of freedom in a unit cell, such that each boundary state localized to $n=1$ has a ``mirror-symmetric" partner localized to $n=N$. In other words, this means that if one finds $p$ boundary state on the boundary $n=1$ ($n=N$) in the case of a lattice with a broken unit cell, one would find $p$ (zero) boundary states on both boundaries in the case of unbroken unit cells. Therefore, for a lattice with no broken unit cells, the value of $\mathcal{P}$ corresponds to the total number of boundary states on either boundary, and $\mathcal{P}=0$ when there are no boundary states. In the following, we always assume that the unit cell is unbroken unless otherwise specified.

\begin{figure}[t]
	\includegraphics[width=\linewidth]{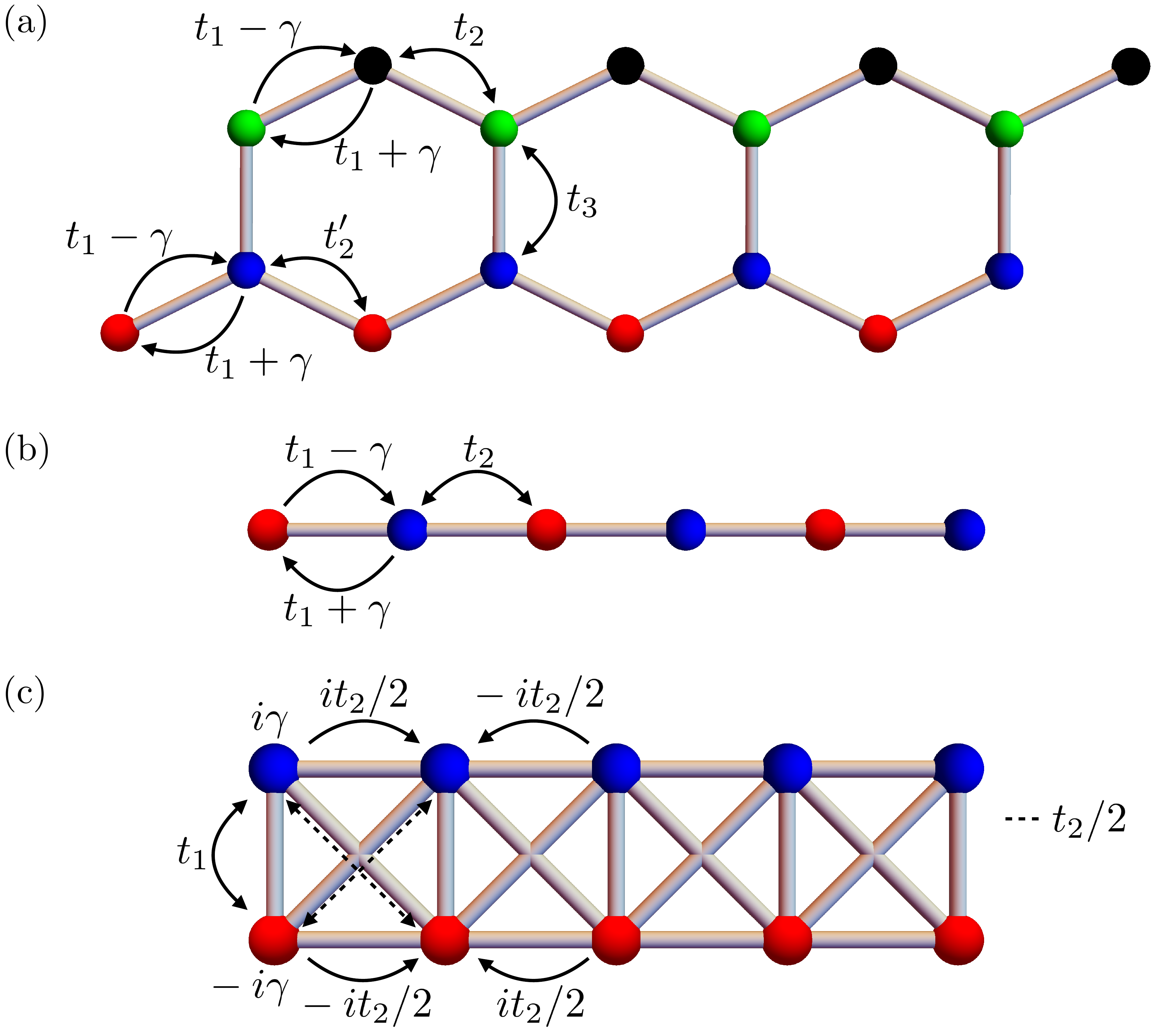}
	\caption{Hamiltonian for (a) the two-leg ladder, (b) the anisotropic SSH chain, and (c) the Lee model. The differently colored sites refer to different sublattice sites.}
	\label{fig:Hamiltonian}
\end{figure}

\begin{figure}
\includegraphics[width=0.8\linewidth]{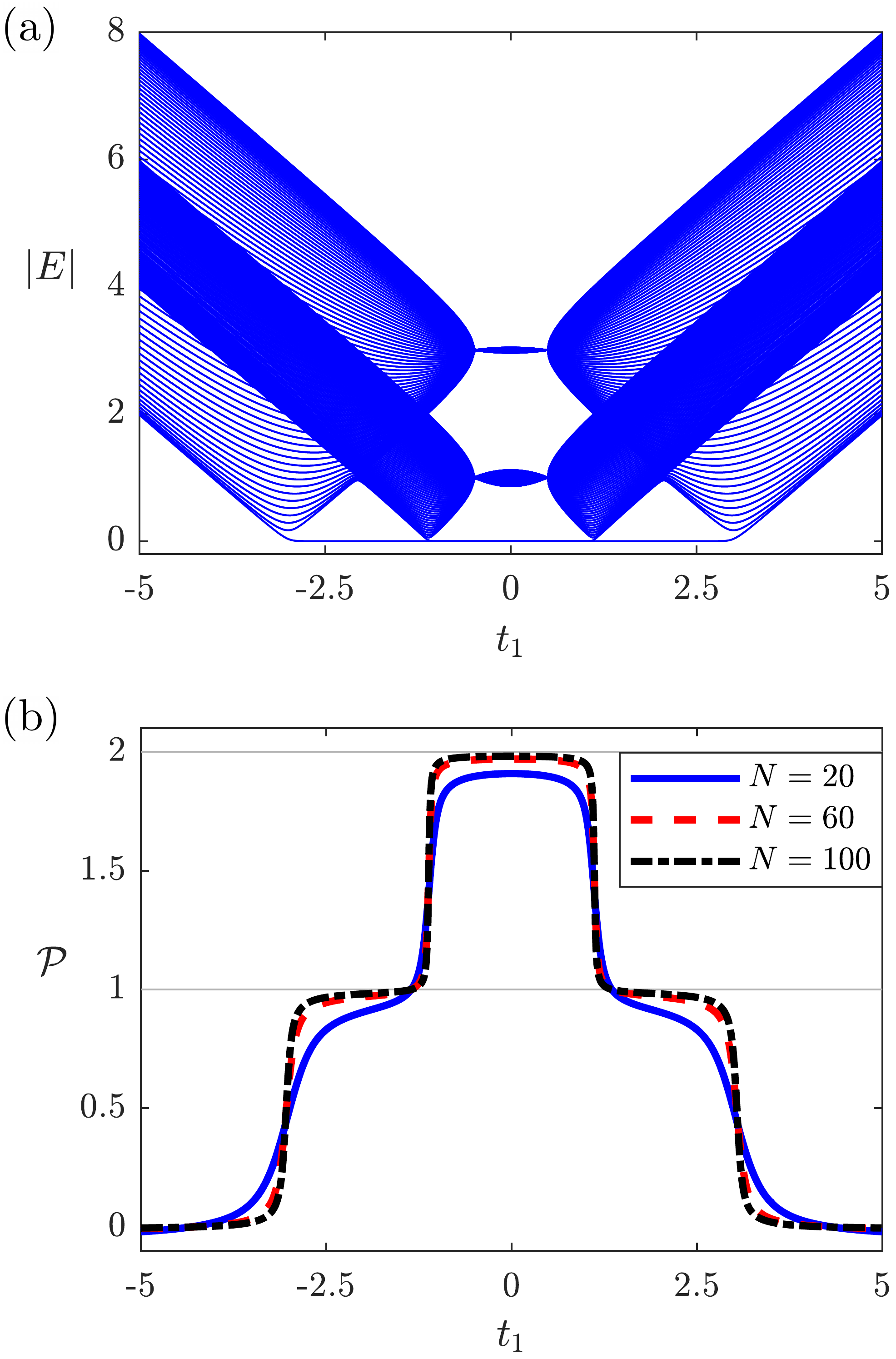}
	\caption{(a) Absolute value of the energy eigenvalues, and (b) the biorthogonal polarization for the two-leg ladder [cf. Fig.~\ref{fig:Hamiltonian}(a)] with $t_2 =3,  t_2' = 1$, $t_3 = 0.1$ and $\gamma= 0.5$. The eigenvalues are computed for $N = 80$ unit cells, and the polarization for the values $N = 20$ (blue solid), $60$ (red dashed) and $100$ (black dashed dotted) unit cells.
	}
	\label{fig:energy_eigenvalues_1}
\end{figure}

To illustrate this in more detail, we consider the example of a two-leg ladder as shown in Fig.~\ref{fig:Hamiltonian}(a). The Hamiltonian reads
\begin{equation}
	H = H_H+H_{AH},
\end{equation}
with
\begin{equation}
\begin{split}
	H_H &= t_1\sum_{n,l}c_{n,l,A}^{\dagger}c_{n,l,B}+t_2\sum_nc_{n+1,a,A}^{\dagger}c_{n,a,B} \\
	&+t_2'\sum_{n}c_{n+1,b,A}^{\dagger}c_{n,b,B}+t_3\sum_{n} c_{n,a,A}^{\dagger}c_{n,b,B}+h.c.,
\end{split}
\end{equation} 
and
\begin{equation}
	H_{AH} = \gamma\sum_{n,l}\left[ c_{n,l,A}^{\dagger}c_{n,l,B}-h.c.\right], 
\end{equation}
where $c_{n,l,\alpha}^{\dagger}$ ($c_{n,l,\alpha}$) creates (annihilates) a state at sublattice $\alpha \in \{A,B\}$ in unit cell $n$ with channel $l = a,b$. When $t_3 = 0$, we obtain two decoupled anisotropic SSH chains [cf. Fig.~\ref{fig:Hamiltonian}(b)]. We note that $H_H$ is Hermitian and that $H_{AH}$ is anti-Hermitian, such that the Hamiltonian $H$ is non-Hermitian. As each individual SSH chain may host at most one end mode at each end, we may find two, one or zero end modes on each of the boundaries of the two-leg ladder depending on the choice of parameters. In Figs.~\ref{fig:energy_eigenvalues_1}(a) and \ref{fig:energy_eigenvalues_1}(b), we plot the band spectrum, and the biorthogonal polarization, respectively, for different systems sizes. We see that the number of end modes varies as a function of $t_1$, where $\mathcal{P}$ [cf. Fig.~\ref{fig:energy_eigenvalues_1}(b)] accurately predicts the number of zero-energy end states in accordance with Fig.~\ref{fig:energy_eigenvalues_1}(a). Indeed, the biorthogonal polarization jumps at those values of $t_1$ at which the band gap closes. We note that the biorthogonal polarization approaches a step function as we increase the system size as advertised.

\subsection{Properties of the biorthogonal polarization}

Here we discuss several interesting properties of the biorthogonal polarization. Firstly, $\mathcal{P}$ is gauge invariant: Assuming that there is an invertible matrix $V$ such that 
\begin{equation}
	\ket{\psi_{\alpha R}} = \sum_{\alpha'} \ket{\phi_{\alpha' R}} V_{\alpha'\alpha},
\end{equation} 
and that both $\phi$ and $\psi$ are normalized according to 
\begin{equation}
	\braket{\psi_{\alpha L}|\psi_{\beta R}} = \braket{\phi_{\alpha L}|\phi_{\beta R}} = \delta_{\alpha\beta}, 
\end{equation}
the corresponding left eigenvector can be written as 
\begin{equation}
	\bra{\psi_{\alpha L}} = \sum_{\alpha'}V^{-1}_{\alpha\alpha'}\bra{\phi_{\alpha'L}}.
\end{equation} 
where $V^{-1}_{\alpha\alpha'} \equiv [V^{-1}]_{\alpha\alpha'}$.
This gives
\begin{equation}
\begin{split}
	&\frac{1}{N}\sum_{\alpha= 1}^M\braket{\psi_{\alpha L}|\left(\sum_nn\hat{\Pi}_n\right)|\psi_{\alpha R}} =\\ &\frac{1}{N}\sum_{\alpha,\beta,\gamma= 1}^MV_{\alpha\beta}^{-1}\braket{\phi_{\beta L}|\left(\sum_nn\hat{\Pi}_n\right)|\phi_{\gamma R}}V_{\gamma\alpha} =\\
	&\frac{1}{N}\sum_{\beta= 1}^M\braket{\phi_{\beta L}|\left(\sum_nn\hat{\Pi}_n\right)|\phi_{\beta R}},
\end{split}
\end{equation}
and we thus find that the polarization is invariant under the change of basis. This is an important characteristic of $\mathcal{P}$, because it means that regardless of the choice one makes for representing $\ket{\psi_{\alpha R/L}}$, one would always find the same result for $\mathcal{P}$.

Another interesting property of the biorthogonal polarization is that it is invariant under unitary transformations that are \emph{local}. Indeed, suppose that we have two Hamiltonians $H$ and $\tilde{H}$ that are related via
\begin{equation}
	\tilde{H} = U^{\dagger}_NHU_N,
\end{equation}
where $U_N$ is a unitary operator of the form $U_N = {\1}_N\otimes U$ with ${\1}_N$ the $N$-dimensional identity matrix and $U$ a $d\times d$-matrix, where $d$ is the total number of degrees of freedom in a unit cell in the system described by $H$. Next, suppose $\ket{\psi_{R/L}}$ is the right/left eigenvector of $H$ with eigenvalue $E$, then $U^{\dagger}\ket{\psi_{R/L}}$ is a right/left eigenvector of $\tilde{H}$ with eigenvalue $E$. This means that the biorthogonal polarization $\tilde{\mathcal{P}}$ of $\tilde{H}$ is given by
\begin{equation}
\tilde{\mathcal{P}} = M-\lim_{N\rightarrow\infty}\frac{1}{N}\sum_{\alpha= 1}^{M}\braket{\psi_{L,\alpha}|U\sum_nn\hat{\Pi}_nU^{\dagger}|\psi_{R,\alpha}},
\end{equation}
where $\hat{\Pi}_n $ projects the states onto unit cell $n$, such that it can be written as $\hat{\Pi}_n = J_{n}\otimes {\1}_{d}$, where $J_{n}$ is an $N$-dimensional matrix with zeros everywhere except at position $(n,n)$ where we have a one. Since $U_N$ and $\hat{\Pi}_n$ have the same block-structure, they must commute, and therefore 
\begin{equation}
	\tilde{\mathcal{P}} = \mathcal{P}.
\end{equation}  
The biorthogonal polarization is thus indeed invariant under this type of unitary transformation. To illustrate the implications of this equality, we consider the anisotropic SSH chain \cite{KuEdBuBe2018} and the Lee model \cite{Le2016} shown in Figs.~\ref{fig:Hamiltonian}(b) and \ref{fig:Hamiltonian}(c), respectively. These two models have the following Bloch Hamiltonians
\begin{align}
H_\textrm{Bloch}^\textrm{SSH} &= \left(t_1 + t_2 \cos k , \, t_2 \sin k + i \gamma, \, 0\right) \cdot \boldsymbol\sigma \label{eq:bloch_Hamiltonian_anistropic_SSH}\\
H_\textrm{Bloch}^\textrm{Lee} &= \left(t_1 + t_2 \cos k , \,0,\, t_2 \sin k + i \gamma\right) \cdot \boldsymbol\sigma, \nonumber
\end{align}
where $\boldsymbol{\sigma}$ is the vector of Pauli matrices. We immediately see that these Hamiltonians can be related via
\begin{equation}
	H_\textrm{Bloch}^\textrm{Lee} = U^{\dagger}H_\textrm{Bloch}^\textrm{SSH}U,
\end{equation}
where $U$ is given by 
\begin{equation}\label{eq:unitary_transform}
	U = \frac{1}{\sqrt{2}} \begin{pmatrix} 1 & i \\ i &1 \end{pmatrix}.
\end{equation}
It is straightforward to show that the Hamiltonians for the anisotropic SSH chain and the Lee model under OBC are also related by a unitary transformation,
\begin{equation}
H_N^\textrm{Lee} = U_N^{\dagger}H_N^\textrm{SSH}U_N, \label{eq:unitary_relation_SSH_Lee_OBC}
\end{equation}
where $U_N$ is defined as above, and $H_N^\textrm{SSH}$ and $H_N^\textrm{Lee}$ are the OBC Hamiltonians with $N$ unit cells for the NH SSH and Lee model, respectively.
Therefore, the biorthogonal polarization of these systems is equivalent. A complementary consequence of the relation in Eq.~(\ref{eq:unitary_relation_SSH_Lee_OBC}) is that the spectra of $H_N^\textrm{SSH}$ and $H_N^\textrm{Lee}$ are identical, while their eigenstates are equal up to permutations inside the unit cell determined by $U$.
Therefore, the exact solutions for the zero-energy end states of the anisotropic SSH chain found in Ref.~\onlinecite{KuEdBuBe2018}, which are discussed in more detail in the next section, are also relevant for Lee's model.

\section{Bulk states and gap closings} \label{sect:bulk_states_gap_closings}

In this section, we study the anistropic SSH chain in more detail through explicit analytical solutions. It was previously shown that this model breaks conventional BBC, and thus displays a spectral instability as well as the non-Hermitian skin effect \cite{KuEdBuBe2018, YaWa2018}. Here, we make use of analytical solutions to study the gap closings of the spectrum in more detail. Additionally, we show that the method developed in Ref.~\onlinecite{KuMiBe2019} for finding all bulk states in the presence of a spectral mirror symmetry can be generalized to this NH model.

\subsection{Closing of the energy gap}
We consider the Bloch Hamiltonian for the anisotropic SSH chain in Eq.~(\ref{eq:bloch_Hamiltonian_anistropic_SSH}) with the energy eigenvalues
\begin{equation}
E_{\pm}^\textrm{PBC}(k) = \pm\sqrt{t_1^2+t_2^2-\gamma^2+2t_1t_2\cos(k)+2\mathrm{i}t_2\gamma\sin(k)}, \label{eq:anistropic_SSH_eigenvalues_PBC}
\end{equation}
and (unnormalized) eigenstates
\begin{align}
\psi_{R,\pm}(k) &=\begin{pmatrix}
t_1+\gamma+t_2e^{-\mathrm{i}k} \\ E_{\pm}^\textrm{PBC}(k)
\end{pmatrix}, \label{eq:right_wavefct_bloch_ssh}\\
\psi_{L,\pm}(k) &=\begin{pmatrix}
t_1-\gamma+t_2e^{-\mathrm{i}k} \\ \left[ E_{\pm}^\textrm{PBC}(k)\right]^* 
\end{pmatrix}.
\end{align}
Here we include the label $\textrm{PBC}$ for the eigenvalues indicating that if we parametrize $k$ such that $k = 2\pi j/N$ for $j = 0, 1,\dots,N-1$ in $E_{\pm}^\textrm{PBC}(k)$, we find the band spectrum for the model with PBCs.

\begin{figure}[b]
	\includegraphics[width=0.9\linewidth]{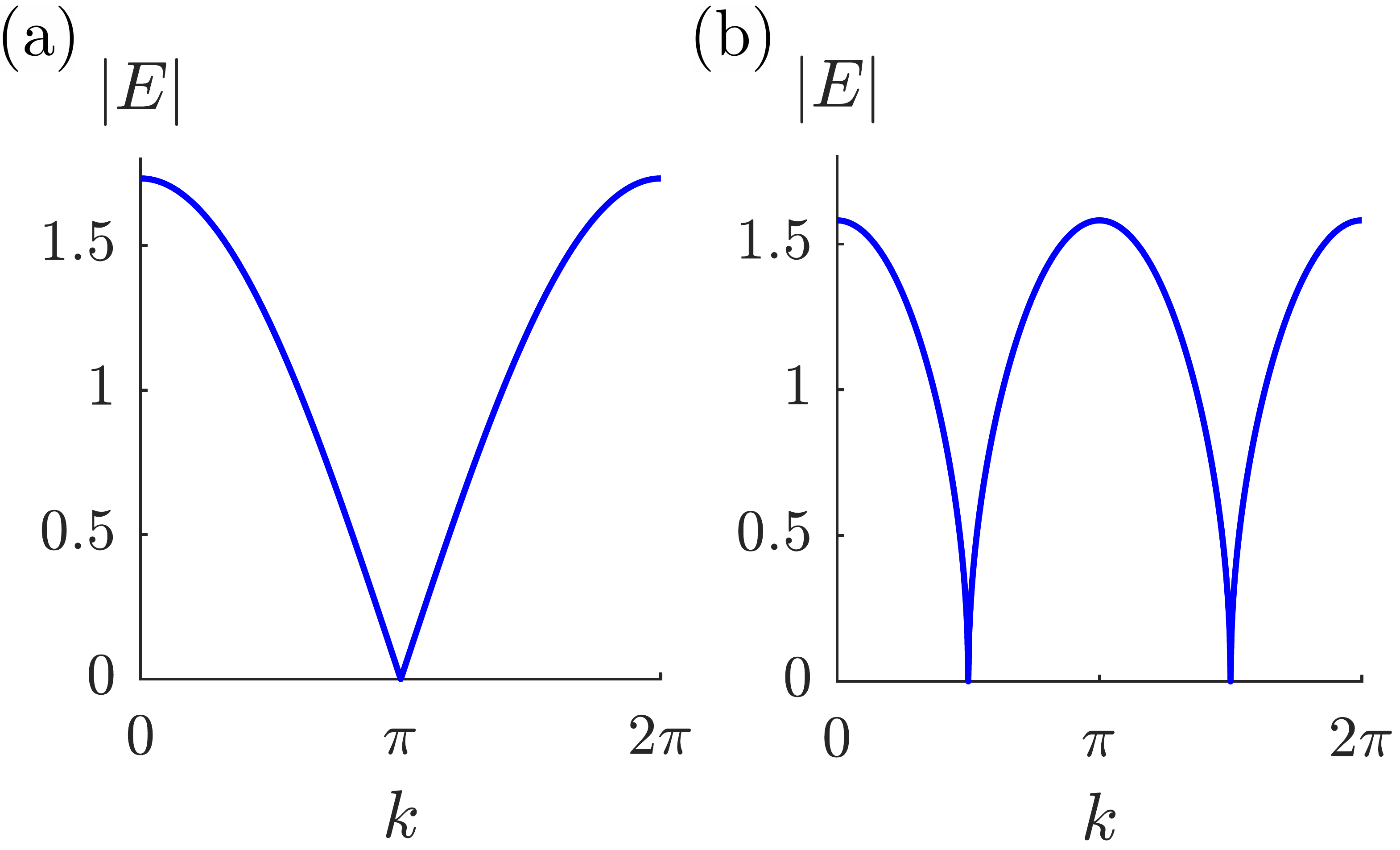}
	\caption{The lowest bulk energy bands of the OBC spectrum [cf. Eq.~(\ref{eq:Eobc})] with a broken unit cell after taking the absolute value for $t_1 = 1$, and (a) $t_2 = \sqrt{3}/2$ and $\gamma=0.5$, and (b) $t_2 = \sqrt{5}/2$ and $\gamma=1.5$. The former case corresponds to the spectrum being real, while the latter corresponds to the spectrum being complex. Notably, the spectrum in (b) also exhibits the genuinely non-Hermitian feature of a non-analytic (square root) dispersion.}
	\label{fig:gap_closings_1}
\end{figure} 

Taking OBCs with a broken unit cell at one boundary, we find that one zero-energy end state appears. In Ref.~\onlinecite{KuEdBuBe2018}, it is shown that this end state is captured by the exact solution
\begin{equation}
	\ket{\psi_{R/L}} = \mathcal{N}_{R/L}\sum_{n= 1}^Nr_{R/L}^nc_{n,A}^{\dagger}\ket{0}, \label{eq:exact_zero_state_sol}
\end{equation}
with $\mathcal{N}_{R/L}$ the normalization, and $r_{R/L} = (t_1\mp\gamma)/t_2$. This end state is delocalized when $|r_Rr_L| = 1$ \cite{KuEdBuBe2018}, which can also be seen from the biorthogonal polarization in Eq.~(\ref{eq:original_biorth_pol}), where $\mathcal{P}$ changes values when $|r_Rr_L| = 1$. Therefore, we expect the bulk-band gap of the OBC spectrum to close at this point. To see this, we compute the eigenvalues of the OBC system from the PBC spectrum by applying a shift in $k$ in the latter, similar to the shift in Refs.~\onlinecite{YaWa2018, KuDw2019}, i.e.,
\begin{equation}
k \rightarrow k- i \log\left(\frac{\sqrt{t_1-\gamma}}{\sqrt{t_1+\gamma}}\right), \label{eq:PBC_to_OBC_shift_in_k}
\end{equation}
such that the OBC spectrum reads
\begin{equation}
\label{eq:Eobc}
\begin{split}
&E_{\pm}^\textrm{OBC}(k) = E_{\pm}^\textrm{PBC}\left[k- i \log\left(\frac{\sqrt{t_1-\gamma}}{\sqrt{t_1+\gamma}}\right) \right] =\\
&\pm\sqrt{t_1^2+t_2^2-\gamma^2+2t_2\sqrt{t_1-\gamma}\sqrt{t_1+\gamma}\cos(k)}.
\end{split}
\end{equation}
We note that $|r_Rr_L| = 1$ yields $t_2 = \pm\sqrt{t_1^2-\gamma^2}$ for $|t_1|>|\gamma|$, such that $E_{\pm}^\textrm{OBC}(k) = 0$ for $k = 0, \pi$ depending on the sign in front of the square root. Similarly, when $|t_1|<|\gamma|$, $|r_Rr_L| = 1$ yields $t_2 = \pm\sqrt{\gamma^2-t_1^2}$, and we find $E_{\pm}^\textrm{OBC}(\pi/2) = E_{\pm}^\textrm{OBC}(3\pi/2) = 0$. This means that the gap closes at these parameters for $k = 0, \pi$ or $k =\pi/2$ and $ 3\pi/2$, respectively, as illustrated in Fig.~\ref{fig:gap_closings_1}. Furthermore, we see that the PBC spectrum $E_{\pm}^\textrm{PBC}(k)$ remains gapped for all $k$ when $|r_Rr_L| =1$. This is in full agreement with the pervious statement that this model features a spectral instability \cite{KuEdBuBe2018}.

\begin{figure}
	\includegraphics[width=0.9\linewidth]{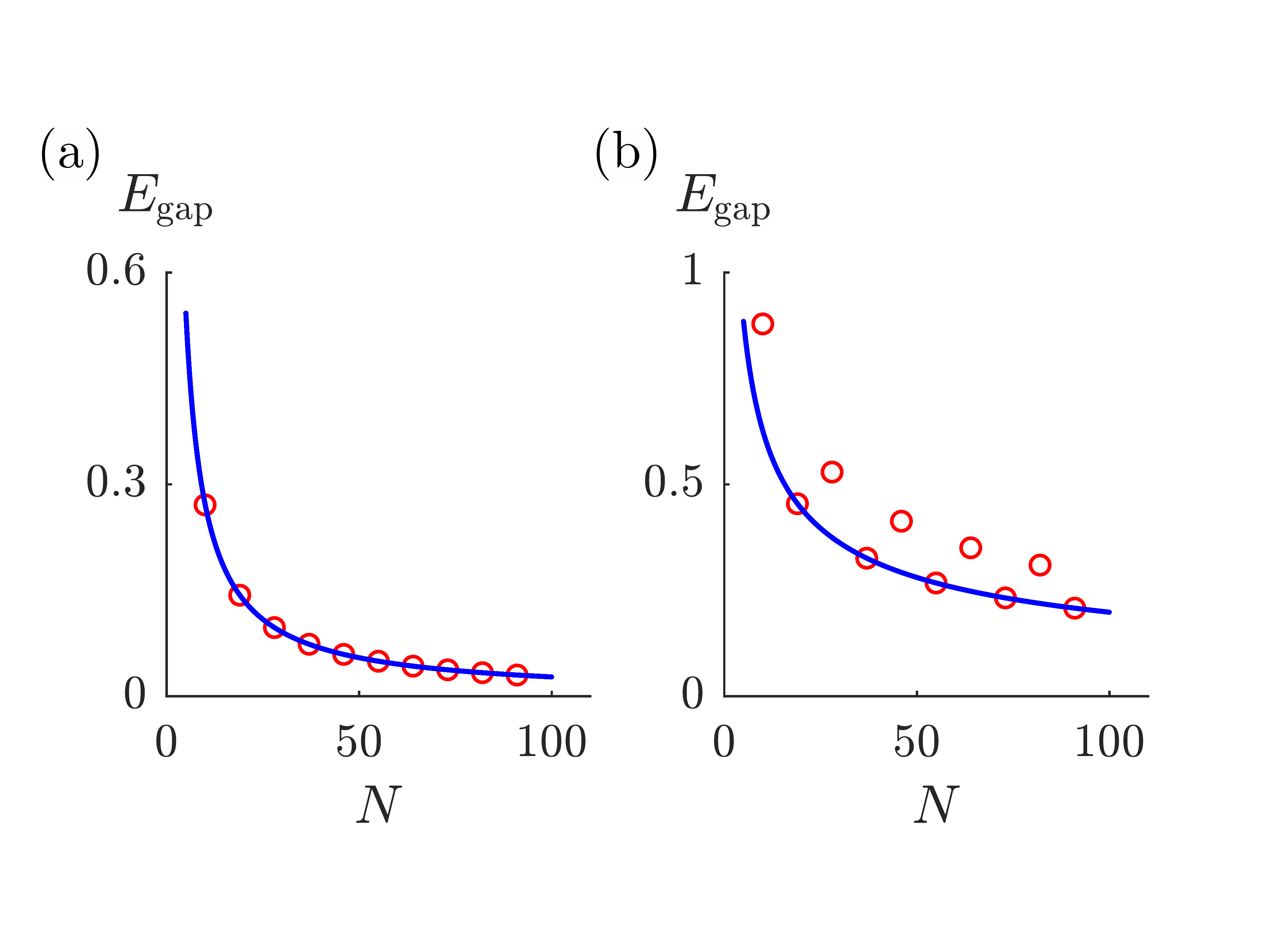}
	\caption{Energy gap closing $E_\textrm{gap}$, which corresponds to the lowest non-zero energy bulk band in the absolute value spectrum, in the OBC spectrum with a broken unit cell as a function of $N$ for (a) $t_2^2 = t_1^2-\gamma^2$, and (b) $t_2^2 = \gamma^2-t_1^2$. The red dots are analytically computed from Eq.~(\ref{eq:Eobc}), whereas the blue lines correspond to the lines $E_\textrm{gap} =  t_2\pi/N$ and $E_\textrm{gap} = t_2\sqrt{\pi/N}$ in the left and right panels, respectively. We see that there is a good agreement between the actual gap sizes (red dots), and the approximated result (blue line).}
	\label{fig:gap_closings_2}
\end{figure}

Performing a series expansion of $E_{\pm}^\textrm{OBC}(k)$ at $t_2 = \pm\sqrt{t_1^2-\gamma^2}$ and $t_2 = \pm\sqrt{\gamma^2-t_1^2}$ around the points $k = \pi$ and $k = \pi/2$, respectively, we find that the gap closes as $E_\textrm{gap}\sim 1/N$ and $E_\textrm{gap}\sim1/\sqrt{N}$, respectively (cf. the blue lines in Fig.~\ref{fig:gap_closings_2}), where $E_\textrm{gap}$ is determined by first taking the absolute value of the energy spectrum and subsequently computing the smallest energy above zero. The latter result is particularly interesting as this type of scaling typically does not occur in Hermitian systems, and in this case it happens when the eigenvalues are complex. We can understand this difference between Hermitian and NH systems by noting that in a Hermitian system the bulk energies are essentially the same under PBC and OBC up to possible boundary states. Under PBC, the energies are periodic functions of $k\sim 1/N$, and have Fourier expansions in $k$. This means that the gap closes at least as fast as $1/N$. This argument clearly fails in the NH case in the absence of a conventional BBC.

\subsection{Exact bulk-state solutions}

In Ref.~\onlinecite{KuMiBe2019}, it was shown by some of the authors of this paper that it is possible to find all bulk-state solutions for a large family of $d$-dimensional lattice models with OBC when the spectrum has a mirror symmetry, i.e., $E(k_\perp, {\bf k}_\parallel) = E(-k_\perp, {\bf k}_\parallel)$, where $k_\perp$ is the momentum in the direction of the open boundary and ${\bf k}_\parallel$ is the crystal momentum in the periodic directions. Here we propose a generalization of this method to the NH realm by specifically focussing on the anistropic SSH chain.

We start by observing that even though the eigenvalues for the PBC spectrum in Eq.~(\ref{eq:anistropic_SSH_eigenvalues_PBC}) are not symmetric under inversion symmetry, i.e., $E_{\pm}^\textrm{PBC}(k) \neq E_{\pm}^\textrm{PBC}(-k)$, the eigenvalues in the case of OBC do display this spectral symmetry, i.e., $E_{\pm}^\textrm{OBC}(k) = E_{\pm}^\textrm{OBC}(-k)$, and we should thus be able to adopt the method developed in Ref.~\onlinecite{KuMiBe2019} to find the eigenstates also in this NH setting even in the presence of the NH skin effect. We note that such a distinction between the PBC and OBC spectra does not exist in the Hermitian case because the spectra would be essentially identical.

To find the right eigenstates, we start by considering a periodic chain with $2N$ unit cells. From Ref.~\onlinecite{KuMiBe2019}, we know that the state in the $n$th unit cell of a \emph{Hermitian} system reads 
\begin{equation}
	\Psi_{R,{\pm}}(k,n) = e^{i k n}\psi_{R,\pm}(k),
\end{equation}
where $\psi_{R,\pm}(k)$ is the eigenstate of the corresponding Bloch Hamiltonian. We now wish to apply the same idea for NH systems. Previously, we saw that to obtain correct results in the OBC case from the PBC solutions, we need to apply a shift in $k$ [cf. Eq.~(\ref{eq:PBC_to_OBC_shift_in_k})]. Applying the same logic here, we make the following ansatz for the bulk eigenstates of the anisotropic SSH chain
\begin{equation}
	\tilde{\Psi}_{R,\pm}(k,n) = \frac{(t_1-\gamma)^{n/2}}{(t_1+\gamma)^{n/2}} e^{\mathrm{i}kn}\tilde{\psi}_{R,\pm}(k),
\end{equation}
where
\begin{equation}
	\tilde{\psi}_{R,\pm}(k) = \begin{pmatrix}
	t_1+\gamma+t_2\dfrac{\sqrt{t_1+\gamma}}{\sqrt{t_1-\gamma}}e^{-\mathrm{i}k} \\ \\ E_{\pm}^\textrm{OBC}(k)
	\end{pmatrix},
\end{equation}
which we obtain from applying the shift in $k$ in Eq.~(\ref{eq:right_wavefct_bloch_ssh}).

\begin{figure}[t]
	\includegraphics[scale=0.7]{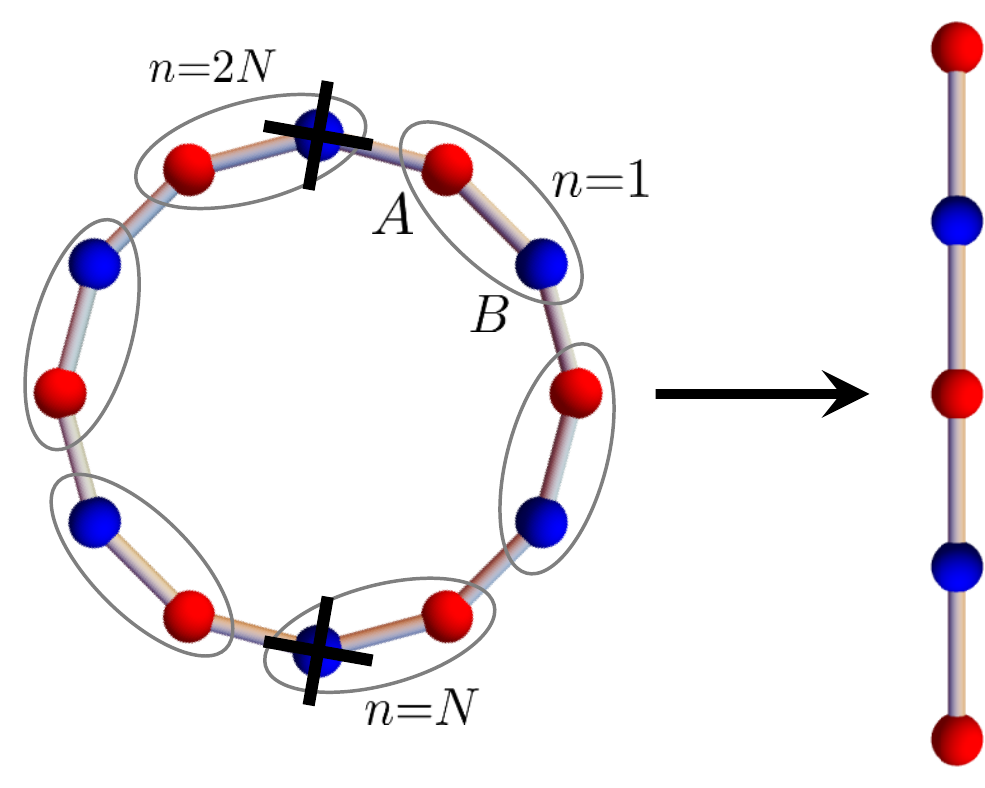}
	\caption{Schematic picture of the periodic chain with $A$ and $B$ sublattices in red and blue, respectively, and the unit cells labelled by $n$ depicted with gray ovals. Upon removing the $B$ sublattices in unit cells $n=0$ ($n = 2N$) and $n = N$, the periodic chain reduces to two open chains with $A$ sublattices at their ends.}
	\label{fig:bulk_state_derivation}
\end{figure}

Next, we assume that all states have a zero amplitude on the $B$ sublattices $n = 0$ (or equivalently $n=2N$) and $n = N$ as shown in Fig.~\ref{fig:bulk_state_derivation}. Upon cutting the chain open by removing the $B$ sites at $n=0$ ($n=2N$) and $n=N$, we end up with two chains with $N$ unit cells. In the following, we focus on one chain, and imagine that we reattach the $B$ sites at both ends. Using the spectral mirror symmetry of $E_{\pm}^\textrm{OBC}(k) = E_{\pm}^\textrm{OBC}(-k)$, we may write the bulk state in the $n$th unit cell as a superposition of $\tilde{\Psi}_{R,\pm}(k,n)$ and $\tilde{\Psi}_{R,\pm}(-k,n)$
\begin{equation}
\begin{split}
	&\Psi_{R,\mathrm{Bulk},\pm}(k,n) = C_1\tilde{\Psi}_{R,\pm}(k,n)+C_2\tilde{\Psi}_{R,\pm}(-k,n)=\\
	& \frac{(t_1-\gamma)^{n/2}}{(t_1+\gamma)^{n/2}}\left[C_1e^{i kn}\tilde{\psi}_{R,\pm}(k)+C_2e^{-ikn}\tilde{\psi}_{R,\pm}(-k)\right],   
\end{split}
\end{equation}
where $k = \pi j /N$ with $j = 1,2, \ldots, N-1$, and impose the boundary condition 
\begin{equation}
	\Psi_{R,\mathrm{Bulk},\pm,B}(k,0) = \Psi_{R,\mathrm{Bulk},\pm,B}(k,N) = 0,
\end{equation}
where the label $\alpha$ in $\Psi_{R,\mathrm{Bulk},\pm,\alpha}(k,n)$ refers to the amplitude of $\Psi_{R,\mathrm{Bulk},\pm}(k,n)$ on sublattice $\alpha$.
The boundary condition leads to
\begin{equation}
	\frac{C_2}{C_1} = -\frac{\tilde{\psi}_{R,\pm,B}(k)}{\tilde{\psi}_{R,\pm,B}(-k)} = -1,
\end{equation}
such that
\begin{equation}\label{eq:AandB}
\begin{split}
	&\Psi_{R,\mathrm{Bulk},\pm,\alpha}(k,n) =\\ &\frac{(t_1-\gamma)^{n/2}}{(t_1+\gamma)^{n/2}}\left[e^{i kn}\tilde{\psi}_{R,\pm,\alpha}(k)-e^{-i kn}\tilde{\psi}_{R,\pm,\alpha}(-k)\right].
\end{split}    
\end{equation}
We thus make the following ansatz for the (unnormalized) bulk states
\begin{equation}\label{eq:full_bulk_states}
	\Psi_{R,\mathrm{Bulk},\pm}(k) = \begin{pmatrix}
	\Psi_{R,\mathrm{Bulk},\pm,A}(k,1) \\
	\Psi_{R,\mathrm{Bulk},\pm,B}(k,1)\\
	\Psi_{R,\mathrm{Bulk},\pm,A}(k,2)\\
	\Psi_{R,\mathrm{Bulk},\pm,B}(k,2)\\
	\vdots\\
	\Psi_{R,\mathrm{Bulk},\pm,B}(k,N-1)\\
	\Psi_{R,\mathrm{Bulk},\pm,A}(k,N)\\
	\end{pmatrix},
\end{equation}
and a straightforward computation of the eigenequation, $H^\textrm{OBC}\Psi_{R,\mathrm{Bulk},\pm}(k) = E^\textrm{OBC}_\pm (k) \Psi_{R,\mathrm{Bulk},\pm}(k)$, shows that these are indeed eigenstates with energy $E^\textrm{OBC}_\pm (k)$. The left eigenstates can be found by making use of the fact that the daggered Hamiltonian of the anistropic SSH chain, i.e. $(H^{\textrm{SSH}})^\dagger$, is simply obtained by transforming $\gamma \rightarrow - \gamma$ in the Hamiltonian $H^\textrm{SSH}$, such that these states are found by taking the complex conjugation of the right states and $\gamma \rightarrow - \gamma$ as shown in Appendix~\ref{app:rel_left_and_right_eigs}.

We notice that Eq.~(\ref{eq:AandB}) has two interesting features: Firstly, we see that its weight explicitly depends on the unit-cell label $n$ and the bulk sates are thus all localized to a boundary when $t_1, \gamma \neq 0$, while the states change localization when passing through $t_1 = 0$ and/or $\gamma = 0$.

Secondly, we notice that the bulk states [and also the shift in Eq.~(\ref{eq:PBC_to_OBC_shift_in_k})] are singular when $t_1= \pm\gamma$. From the Hamiltonian in Fig.~\ref{fig:Hamiltonian}(b), we immediately see that at these values of $t_1$ it is possible to hop only in one direction, and consequently all eigenstates are exactly localized at the boundary. Additionally, the bulk spectrum only has two eigenvalues, $\pm t_2$ [cf. Eq.~(\ref{eq:Eobc})]. This behavior of the eigenstates and eigenvalues is associated with the presence of EPs, and indeed, these points correspond to EPs with an order that scales with system size \cite{KuDw2019}. The bulk states thus merge into two different states as they approach one of the EPs.

A natural question is what happens with the bulk-state solution in Eq.~(\ref{eq:AandB}) at these points? We note that an eigenvector of an operator is only determined up to multiplication by a scalar, and we show in Appendix~\ref{app:bulk_states_at_eps} that we can choose a scalar in such a way that the limit $t_1 \rightarrow \pm \gamma$ exists. We note these multiplicative factors differs for the two EPs. We then find that the right and left eigenstates only have nonzero amplitudes at opposite boundaries [e.g., Eqs.~(\ref{eq:right_wave_fct_A_at_eps}), (\ref{eq:right_wave_fct_B_at_eps}), (\ref{eq:left_wave_fct_A_at_eps}) and (\ref{eq:left_wave_fct_B_at_eps})], which means that the states are not normalizable, which is consistent with the behavior of an EP.

\subsection{Twisted states}

Let us now study what happens with the bulk states at the gap closings in the spectrum, i.e., at $|r_Rr_L| = 1$. To gain further understanding of this, we normalize the bulk states found in the previous section for $|t_1|>|\gamma|$ and $t_1+\gamma>0$ at the point $t_2 = \sqrt{t_1^2-\gamma^2}$,
\begin{equation}
\braket{\Psi_{L,\mathrm{Bulk},\pm}|\Psi_{R,\mathrm{Bulk},\pm}} = -16N(t_1^2-\gamma^2)\cos^2\left(\frac{k}{2}\right){,} 
\end{equation}
such that the normalized state takes the form
\begin{equation}
	\ket{\Psi_{R,\mathrm{Bulk},\pm}(k)} = \ket{\Psi_{\textrm{twist},A}(k)}\pm\ket{\Psi_{\textrm{twist},B}(k)},
\end{equation}
where 
\begin{equation}
\begin{split}
\ket{\Psi_{\textrm{twist},A}(k)} &= \frac{1}{\sqrt{N}}\sum_{n = 1}^{N}\left( \frac{t_1-\gamma}{t_2}\right)^{n-1}\\& \times \sin\left( \frac{(2n-1)k}{2}\right) c_{n,A}^{\dagger}\ket{0},
\end{split}
\end{equation}
and
\begin{equation}
\ket{\Psi_{\textrm{twist},B}(k)} = \frac{1}{\sqrt{N}}\sum_{n = 1}^{N-1}\left( \frac{t_1-\gamma}{t_2}\right)^{n}\sin\left( nk\right) c_{n,B}^{\dagger}\ket{0}.
\end{equation}
We note these states are \emph{twisted} versions of the eigenstates of the zero-energy mode in Eq.~(\ref{eq:exact_zero_state_sol}) in the sense that while they have the same amplitude, they also feature an additional phase that changes along the chain. Indeed, the larger the length of the chain, the smaller the local twist becomes. This means that in the limit of going to infinite system size, the twisted state has the same energy as the zero-energy state, and they are thus degenerate. Therefore, the bulk-band gap must close when $t_2 = \sqrt{t_1^2-\gamma^2}$. We repeat this calculation for $|t_1|<|\gamma|$ and $\gamma+t_1>0$ at the point  $t_2 = \sqrt{\gamma^2-t_1^2}$ in Appendix~\ref{app:twisted_states}, and show that we arrive at the same conclusion.

\section{Discussion} \label{sect:discussion}
In this paper, we have further expanded the toolbox of methods for characterising and finding analytical solutions in NH systems. Specifically, we generalized the biorthogonal polarization $\mathcal{P}$ first proposed in Ref.~\onlinecite{KuEdBuBe2018} to be applicable to quasi-one-dimensional systems with any number of boundary modes. We showed that $\mathcal{P}$ is gauge invariant, which is a crucial property for measurable quantities. Additionally, we showed that $\mathcal{P}$ is invariant under local unitary transformations. As a consequence, the biorthogonal polarizations for models that are related via the same unitary transformation are thus equivalent. We emphasize that even though we explicitly treated one-dimensional examples to study the generalized biorthogonal polarization, the formalism developed in this paper works for any model with boundaries of codimension one, i.e., for any $d$-dimensional system with boundaries of $d-1$ dimensions. Moreover, we believe that the generalized biorthogonal polarization should be readily further generalizable to systems with boundaries of higher codimension. Indeed, in Ref.~\onlinecite{EdKuBe2019}, the quantity $\braket{\psi_L|\Pi_{n, n',\ldots}|\psi_R}$ with $n,n',\ldots$ labelling the unit cells in the different directions was shown to accurately capture the presence of corner and hinge states for lattice models with OBC in more directions.

We also studied the anistropic SSH chain in Fig.~\ref{fig:Hamiltonian}(b) in great detail. By making use of a shift in $k$ that connects the PBC and OBC cases to each other [cf. Eq.~(\ref{eq:PBC_to_OBC_shift_in_k})], it is possible to find analytical expressions for the eigenvalues as already presented in Ref.~\onlinecite{KuDw2019}. Making use of these solutions for the OBC case, we showed that the finite size gaps may close slowly in NH systems as compared to Hermitian ones. In particular, we find $E_\textrm{gap} \propto N^{-1/p}$, while gap closings in the Brillioun zone may scale as $E_\textrm{gap} \propto (k-k_0)^{1/p}$ with $p$ some integer. These gap closings are sharper than those in Hermitian systems, which remain analytic. While we focus on two-band models in this work, where $p\leq 2$, this type of scaling is expected to persist once more bands are considered. In this context we note, however, that two-band models are sufficient to study (ordinary) band crossings as these generically occur in a three-dimensional parameter space, and that models with band crossings involving more bands generically require the tuning of a large number of parameters.

By extending the method in Ref.~\onlinecite{KuMiBe2019} and making use of the shift in Eq.~(\ref{eq:PBC_to_OBC_shift_in_k}), we were able to not only find the eigenvalues but also to find closed form analytical equations for all bulk states for the anisotropic SSH model with OBC in addition to the end state solutions that were already found in Ref.~\onlinecite{KuEdBuBe2018}. Making use of these solutions, we were able to prove that the band gap indeed closes when $|r_R r_L|=1$ \cite{KuMiBe2019}. While we here only showed that this method works for a specific example, the anisotropic SSH chain, we believe that it should be applicable to a large family of quasi-one-dimensional NH lattice models with $E^\textrm{OBC}(k_\perp, {\bf k}_\parallel) = E^\textrm{OBC}(-k_\perp, {\bf k}_\parallel)$ in analogy to the Hermitian version of this approach \cite{KuMiBe2019}.

Our analytical results complement a large body of recent numerical and experimental studies on non-Hermitian systems and offer as such complementary and detailed insights into an active field of contemporary and cross-disciplinary physics.

\acknowledgments{
EE, FKK and EJB acknowledge useful discussions with Jan Budich and Guido van Miert on related projects. 
We used the Mathlab extension \texttt{ADVANPIX} to produce the plot for the polarization. 
E.E. and E.J.B. are funded by the Swedish Research Council (VR) and the Knut and Alice Wallenberg Foundation. 
F.K.K. is supported by the Max Planck Institute of Quantum Optics (MPQ) and the Max-Planck-Harvard Research Center for Quantum Optics (MPHQ).  
T.Y. was supported in part by JSPS KAKENHI Grant No.~JP20H04627.
} 

\appendix

\section{Relation between the right and left eigenstates} \label{app:rel_left_and_right_eigs}

Here we show that in a system with open boundary conditions, we can find the left eigenstates from the right eigenstates by using that the Hamiltonian is symmetric under Hermitian conjugation and $\gamma\mapsto-\gamma$ respectively.
Therefore, the left eigenstates are given by the complex conjugated right eigenstates with mirrored $\gamma$. To see this, suppose that 
\begin{equation}
	H(\gamma)\ket{\psi_R^n(\gamma)} = E_n(\gamma)\ket{\psi_R^n(\gamma)}.
\end{equation}  
We wish to find the corresponding left eigenstate $\ket{\psi_L^n(\gamma)}$, which we know satiesfies
\begin{equation}
	H^{\dagger}(\gamma)\ket{\psi_{L}^n(\gamma)} = E_n^*\ket{\psi_L^n(\gamma)}.
\end{equation}
Complex conjugating this equation, letting $\gamma\mapsto-\gamma$ and using that we know that $H^{\dagger}(-\gamma) = H^{T}(-\gamma) = H(\gamma)$ and that Eq. \eqref{eq:Eobc}, gives us $E_n(-\gamma) = E_n(\gamma)$, we get
\begin{equation}
	H(\gamma)\ket{\psi_{L}^n(-\gamma)}^* = E_n(\gamma)\ket{\psi_L^{n}(-\gamma)}^*,
\end{equation} 
and we get
\begin{equation}
	\ket{\psi_L^n(\gamma)} = \ket{\psi_R^n(-\gamma)}^*. \label{eq:relation_right_and_left_eigs}
\end{equation}

For the Bloch Hamiltonian, a slightly different argument must be used as it not symmetric under Hermitian conjugation and $\gamma\mapsto-\gamma$ separately, but rather under the composition of those. For periodic boundary conditions, we also have $E(k,\gamma) = E^*(k,-\gamma)$, in contrast to the case for open boundary conditions. The equation for the left eigenstates of the Bloch Hamiltonian is given by
\begin{equation}
H^{\dagger}_{\mathrm{Bloch}}(k,\gamma)\ket{\psi_{L}(k,\gamma)} = E^*(k,\gamma)\ket{\psi_L(k,\gamma)}.
\end{equation}
Letting $\gamma\mapsto-\gamma$, we get
\begin{equation}
H^{\dagger}_{\mathrm{Bloch}}(k,-\gamma)\ket{\psi_{L}(k,-\gamma)} = E^*(k,-\gamma)\ket{\psi_L(k,-\gamma)},
\end{equation}
which implies
\begin{equation}
H_{\mathrm{Bloch}}(k,\gamma)\ket{\psi_{L}(k,-\gamma)} = E(k,\gamma)\ket{\psi_L(k,-\gamma)},
\end{equation}
such that
\begin{equation}
	\ket{\psi_{L}(k,\gamma)} = \ket{\psi_{R}(k,-\gamma)}.
\end{equation}

\section{Bulk states at the exceptional points} \label{app:bulk_states_at_eps}

We study what happens when we approach the exceptional point at $t_1 = \gamma>0$ from $t_1>\gamma$. We see that if the right eigenstates in Eq.~(\ref{eq:AandB}) are multiplied by 
\begin{equation}
	\frac{\sqrt{t_1+\gamma}}{\sqrt{t_1-\gamma}}\frac{1}{\sin(k)},
\end{equation} 
and the left eigenstates by
\begin{equation}
	\frac{(t_1-\gamma)^{(N-1)/2}}{(t_1+\gamma)^{(N-1)/2}}\frac{1}{\sin(k(N-1))},
\end{equation}
we find
\begin{equation}
\begin{split}
	& \Psi_{R,\mathrm{Bulk},\pm,A}(k,n)= 2 i \frac{(t_1-\gamma)^{(n-1)/2}}{(t_1+\gamma)^{(n-1)/2}}\\& \times \left[(t_1+\gamma)\frac{\sin(kn)}{\sin(k)}+ t_2\frac{\sqrt{t_1+\gamma}}{\sqrt{t_1-\gamma}}\frac{\sin(k(n-1))}{\sin(k)} \right],
\end{split} 
\end{equation}
and
\begin{equation}
\Psi_{R,\mathrm{Bulk},\pm,B}(k,n) = 2 i \frac{(t_1-\gamma)^{(n-1)/2}}{(t_1+\gamma)^{(n-1)/2}}E_{\pm}^\textrm{OBC}(k)\frac{\sin(kn)}{\sin(k)}. 
\end{equation}
We see that
\begin{equation}
	\lim_{t_1\rightarrow\gamma}\Psi_{R,\mathrm{Bulk},\pm,A}(k,n) = \begin{cases} 4 i \gamma & \mbox{if } n = 1,\\ 2 i t_2 & \mbox{if } n = 2{,} \\ 0 & \mbox{if } n>2, \end{cases} \label{eq:right_wave_fct_A_at_eps}
\end{equation}
and
\begin{equation}
\lim_{t_1\rightarrow\gamma}\Psi_{R,\mathrm{Bulk},\pm,B}(k,n) = \begin{cases} \pm 2 i t_2 & \mbox{if } n = 1, \\ 0 & \mbox{if } n>1. \end{cases} \label{eq:right_wave_fct_B_at_eps}
\end{equation}
In a similar fashion, the corresponding left eigenstates, after multiplication by the appropriate factor, are given by 
\begin{equation}
\begin{split}
\Psi_{L,\mathrm{Bulk},\pm,A}(k,n) = -2 i \frac{(t_1-\gamma)^{(N-n-1)/2}}{(t_1+\gamma)^{(N-n-1)/2}}\\ \left[(t_1-\gamma)\frac{\sin(kn)}{\sin(k(N-1))}+t_2\frac{\sqrt{t_1-\gamma}}{\sqrt{t_1+\gamma}}\frac{\sin(k(n-1))}{\sin(k(N-1))} \right], 
\end{split}
\end{equation}
and
\begin{equation}
\begin{split}
\Psi_{L,\mathrm{Bulk},\pm,B}(k,n) =& -2 i \frac{(t_1-\gamma)^{(N-n-1)/2}}{(t_1+\gamma)^{(N-n-1)/2}}\\&E_{\pm}^\textrm{OBC}(k)\frac{\sin(kn)}{\sin(k(N-1))}, 
\end{split}
\end{equation}
and we get
\begin{equation}
\lim_{t_1\rightarrow\gamma}\Psi_{L,\mathrm{Bulk},\pm,A}(k,n) = \begin{cases} -2 i t_2 & \mbox{if } n = N, \\ 0 & \mbox{if } n<N, \end{cases} \label{eq:left_wave_fct_A_at_eps}
\end{equation}
and
\begin{equation}
\lim_{t_1\rightarrow\gamma}\Psi_{L,\mathrm{Bulk},\pm,B}(k,n) = \begin{cases} \mp 2 i t_2 & \mbox{if } n = N-1, \\ 0 & \mbox{if } n<N-1. \end{cases} \label{eq:left_wave_fct_B_at_eps}
\end{equation}
We note that we need to pick different prefactors for the eigenstates in order for them to approach the correct states at the other EP at $t_1 = -\gamma$ but that a similar solution would be found with the difference that the right and left eigenstates are now localized to the opposite boundaries.

\section{Twisted states} \label{app:twisted_states}

For $|\gamma|>|t_1|$ and $t_1+\gamma>0$ at the point $t_2 = \sqrt{\gamma^2-t_1^2}$, we instead get
\begin{equation}
\braket{\Psi_{L,\mathrm{Bulk},\pm}|\Psi_{R,\mathrm{Bulk},\pm}} =8 i N(t_1^2-\gamma^2)\cos(k).
\end{equation}
We again get

\begin{equation}
\ket{\Psi_{R,\mathrm{Bulk},\pm}(k)} = \ket{\Psi_{\textrm{twist},A}(k)}\pm\ket{\Psi_{\textrm{twist},B}(k)},
\end{equation}
but in this case, we have
\begin{equation}
\begin{split}
\ket{\Psi_{\textrm{twist},A}(k)} &= -\frac{1}{\sqrt{N}}\frac{1}{\sqrt{-2\mathrm{i}\cos(k)}}\\& \times \sum_{n = 1}^{N}\frac{1}{i^{n-1}}\left( \frac{t_1-\gamma}{t_2}\right)^{n-1}\left[\sin(kn)\right.\\&\left. -i\sin(k(n-1))\right] c_{n,A}^{\dagger}\ket{0},
\end{split}
\end{equation}
and
\begin{equation}
\begin{split}
\ket{\Psi_{\textrm{twist},B}(k)} &=\\ -\frac{1}{\sqrt{N}}\frac{|\cos(k)|}{\cos(k)}&\sum_{n = 1}^{N-1}(-i)^n\left( \frac{t_1-\gamma}{t_2}\right)^{n}\sin\left( nk\right) c_{n,B}^{\dagger}\ket{0}.
\end{split}
\end{equation}

The states are not as nice as in the point $t_2^2 = t_1^2-\gamma^2$, but one can still see that they are sums of trigonometric functions that one can interpret as a slow change of phase.

\end{document}